\documentstyle{mn}

\newif\ifAMStwofonts

\input psfig
\ifoldfss
  \newcommand{\rmn}[1] {{\rm #1}}

  \ifCUPmtlplainloaded \else
    \NewTextAlphabet{textbfit} {cmbxti10} {}
    \NewTextAlphabet{textbfss} {cmssbx10} {}
    \NewMathAlphabet{mathbfit} {cmbxti10} {} 
    \NewMathAlphabet{mathbfss} {cmssbx10} {} 
  \fi
  \ifAMStwofonts
    \ifCUPmtlplainloaded \else
      \NewSymbolFont{upmath} {eurm10}
      \NewSymbolFont{AMSa} {msam10}
      \NewMathSymbol{\upi}     {0}{upmath}{19}
      \NewMathSymbol{\umu}     {0}{upmath}{16}
      \NewMathSymbol{\upartial}{0}{upmath}{40}
      \NewMathSymbol{\leqslant}{3}{AMSa}{36}
      \NewMathSymbol{\geqslant}{3}{AMSa}{3E}

      \let\leq=\leqslant 
       
    \fi
  \fi
\fi 

\ifnfssone
  \newmathalphabet{\mathit}
  \addtoversion{normal}{\mathit}{cmr}{m}{it}
  \addtoversion{bold}{\mathit}{cmr}{bx}{it}
  \newcommand{\rmn}[1] {\mathrm{#1}}

  \newmathalphabet{\mathbfit} 
  \addtoversion{normal}{\mathbfit}{cmr}{bx}{it}
  \addtoversion{bold}{\mathbfit}{cmr}{bx}{it}
  \newmathalphabet{\mathbfss} 
  \addtoversion{normal}{\mathbfss}{cmss}{bx}{n}
  \addtoversion{bold}{\mathbfss}{cmss}{bx}{n}
  \ifAMStwofonts
    \ifCUPmtlplainloaded \else
      \UseAMStwoboldmath
      \makeatletter
      \new@mathgroup\upmath@group
      \define@mathgroup\mv@normal\upmath@group{eur}{m}{n}
      \define@mathgroup\mv@bold\upmath@group{eur}{b}{n}
      \edef\UPM{\hexnumber\upmath@group}
      \new@mathgroup\amsa@group
      \define@mathgroup\mv@normal\amsa@group{msa}{m}{n}
      \define@mathgroup\mv@bold\amsa@group{msa}{m}{n}
      \edef\AMSa{\hexnumber\amsa@group}
      \makeatother
      \mathchardef\upi="0\UPM19
      \mathchardef\umu="0\UPM16
      \mathchardef\upartial="0\UPM40
      \mathchardef\leqslant="3\AMSa36
      \mathchardef\geqslant="3\AMSa3E

      \let\leq=\leqslant 

    \fi
  \fi
\fi 

\ifnfsstwo
  \newcommand{\rmn}[1] {\mathrm{#1}}

  \DeclareMathAlphabet{\mathbfit}{OT1}{cmr}{bx}{it}
  \SetMathAlphabet\mathbfit{bold}{OT1}{cmr}{bx}{it}
  \DeclareMathAlphabet{\mathbfss}{OT1}{cmss}{bx}{n}
  \SetMathAlphabet\mathbfss{bold}{OT1}{cmss}{bx}{n}
  \ifAMStwofonts
    \ifCUPmtlplainloaded \else
      \DeclareSymbolFont{UPM}{U}{eur}{m}{n}
      \SetSymbolFont{UPM}{bold}{U}{eur}{b}{n}
      \DeclareSymbolFont{AMSa}{U}{msa}{m}{n}
      \DeclareMathSymbol{\upi}{0}{UPM}{"19}
      \DeclareMathSymbol{\umu}{0}{UPM}{"16}
      \DeclareMathSymbol{\upartial}{0}{UPM}{"40}
      \DeclareMathSymbol{\leqslant}{3}{AMSa}{"36}
      \DeclareMathSymbol{\geqslant}{3}{AMSa}{"3E}

      \let\leq=\leqslant 

    \fi
  \fi
\fi 

\ifCUPmtlplainloaded \else
  \ifAMStwofonts \else 
    \def\upi{\pi}
    \def\umu{\mu}
    \def\upartial{\partial}
  \fi
\fi

\title{The Outbursts of Dwarf Novae}
\author[M. R. Truss, J. R. Murray, G. A. Wynn and R.G. Edgar]
       {M. R. Truss, J. R. Murray, G. A. Wynn and R.G. Edgar\\
        Department of Physics and Astronomy, University of Leicester,
        University Road, Leicester, LE1 7RH, UK}

\pubyear{2000}

\begin{document}

\maketitle


\begin{abstract}

We present a numerical scheme for the evolution of an accretion disc
through a dwarf nova outburst. We introduce a time-varying artificial
viscosity into an existing smoothed particle hydrodynamics code
optimised for two and three-dimensional simulations of accretion
discs. The technique gives rise to coherent outbursts and can easily be
adapted to include a complete treatment of thermodynamics. We apply a
two-dimensional isothermal scheme to the system SS Cygni and present a
wide range of observationally testable results. 

\end{abstract}

\begin{keywords}

accretion, accretion discs - instabilities - hydrodynamics - methods:
numerical - binaries:close - novae, cataclysmic variables.

\end{keywords}

\section{Introduction}

Dwarf novae are a class of cataclysmic variable which undergo regular
but aperiodic phases lasting several days during which the
system brightness increases by two to four magnitudes. These
are the well known {\em{normal outbursts}} of dwarf novae, and they recur
on time-scales of weeks to months. There is a well-known bimodal
distribution of orbital periods of cataclysmic variables, with a
dearth of systems having orbital periods in the range
2.2 $\leq P_{\rmn{orb}} \leq $ 2.8 hours. This is known as the
period gap. Dwarf novae which lie above the period gap and only display
normal outbursts are classified as U Geminorum systems, after
their template. There is a roughly linear period-mass ratio relation
for interacting binaries (Frank, King and Raine, 1995). The SU Ursae
Majoris systems, which lie below the period gap and have more extreme
mass ratios $q={M_{2}\over{M_{1}}}<0.25$ show longer, slightly brighter
superoutbursts lasting for ten days or more. These occur in
addition to normal outbursts (typically one superoutburst occurs for
every 5 to 15 normal outbursts) and show a superimposed periodic
variation in brightness at supermaximum (superhumps). In this
paper we introduce a numerical method for studying dwarf nova
outbursts and apply it to a system of the U Gem class. In a future
paper we will present the application to the SU UMa class (Truss et al, 2000).
 
Historically, two theories have been put forward to explain
the outbursts of dwarf novae. Paczy\'{n}ski et al (1969)
suggested that a low mass Roche-lobe-filling secondary is potentially
unstable and mass transfer may occur quickly enough that its
convective envelope would become radiative. The enhanced radiation
susequently stabilises the mass transfer and in this way an outburst
cycle is initiated. However, observations of the hot-spot (the bright
point at which material enters the disc) do not show an increase in
brightness during outburst, as would be predicted by such a
mass-ransfer instability model (see for example, Rutten et al (1992)).
The currently favoured model, and the model which we apply here, is that of an instability in the
accretion disc itself. Integration of the density profile normal to
the plane of the disc $\rho(z)$ yields a relationship between surface
density $\Sigma$ and temperature T (or mass-transfer rate) for an
annulus in the disc. 
This is the well-known {\em{S-curve}} and is
shown schematically in Figure~1. It represents the locus of
points for which the annulus remains in thermal equilibrium. The
physical basis which underpins the S-curve is the onset of hydrogen
ionization. Hence the
upper and lower branches of the curve correspond to the high and low
opacity states of ionized and neutral
hydrogen. The opacity function $\kappa(T)$ is very steep in the range
6000 - 7000 K; such an abrupt change leads to the instability. 
We can define the local viscous energy generation rate per unit area by

\begin{equation}
 Q^{+}=\int{q^{+}(z)dz}
\end{equation}
\noindent
where $q^{+}$ is the local viscous heat generation rate per unit area,
assumed to be concentrated in the centre of the disc. Heat losses can
be assumed to be from black-body dissipation from the disc surface. If
the surface temperature is $T_{\rmn{s}}$, then the local energy dissipation
rate is

\begin{equation}
 Q^{-}=2\sigma{T_{\rmn{s}}}^{4},
\end{equation}
\noindent
where each surface of the disc contributes $\sigma{T_{\rmn{s}}}^{4}$.
In thermal equilibrium, $Q^{+}(T_{\rmn{c}})=Q^{-}(T_{\rmn{s}})$, where
$T_{c}$ is the central disc temperature.

\begin{figure}
\psfig{file=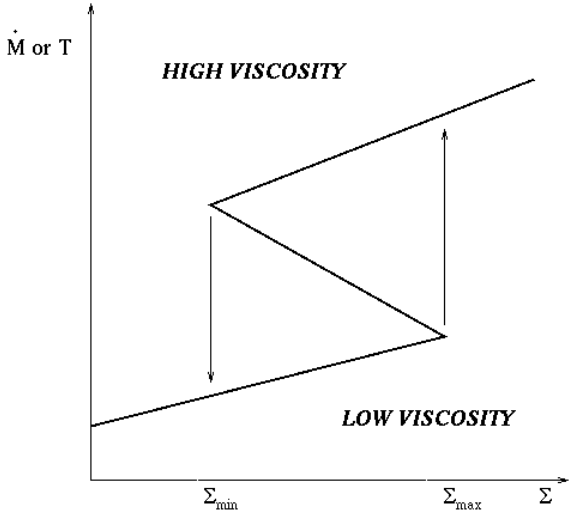,width=8cm}
\caption{The limit-cycle behaviour of an accretion disc in a dwarf
nova.}
\label{scurve}
\end{figure}

An annulus residing on the lower (quiescent) or
upper (outburst) branch of the curve subject to perturbations in local
surface density can evolve on a viscous time-scale and remain in
thermal equilibrium. No stable solution is available to an annulus in
the middle branch, however. Here, if the mass-transfer rate (and hence
$T_{\rmn{c}}$) is increased, $Q^{+}$ increases faster than $Q^{-} $and
$T_{\rmn{c}}$ rises yet further. Therfore, at the inflection points of the
curve, the disc will evolve on a thermal time-scale (that is, very
rapidly) to the opposite branch. In this way a limit cycle of
quiescence and outburst is established. We interpret the upper and
lower branches of the curve as high and low viscosity states. The critical surface densities
at the points of inflection have been calculated by Cannizzo et al
(1988) :

\begin{equation}
 \Sigma_{\rmn{max}} =
11.4R_{10}^{1.05}M_{1}^{-0.35}\alpha_{\rmn{cold}}^{-0.86} ~\rm{gcm^{-2}}
\end{equation}
\begin{equation}
\Sigma_{\rmn{min}} =
8.25R_{10}^{1.05}M_{1}^{-0.35}\alpha_{\rmn{hot}}^{-0.8} ~\rm{gcm^{-2}} 
\end{equation}
\noindent
where $R_{10}$ is the radius in units of $10^{10}$cm, $M_{1}$ is the mass of
the primary in solar masses and $\alpha_{\rmn{hot}}$ and
$\alpha_{\rmn{cold}}$ are the Shakura-Sunyaev viscosity parameters in
the high and low states.
Note that these are very close to linear radial dependences.
An excellent review of the dwarf nova disc instability is given by
Cannizzo (1993a).

We begin with a full explanation of our model and the numerical
techniques involved in applying a smoothed particle hydrodynamics
(SPH) approach to the problem. In Section 3 we demonstrate that our
model is physically consistent with a real dwarf nova and present a
wide range of simulated observables.

\section{Modelling Dwarf Nova Outbursts}

\subsection{The Smooth Particle Hydrodynamics Technique}

In this section we discuss substantial developments made to an
existing smooth particle hydrodynamics accretion disc code that
enable us to model the complete outburst cycle of a dwarf nova.
The principal change is the modification of the dissipation term to
make it a 
function of local disc conditions.
Smooth particle hydrodynamics (SPH) is a Lagrangian method for modelling the
dynamics of fluids. A continuous medium is modelled by a
collection of particles that each move with the local fluid
velocity.
Fluid properties at any given point are determined by
interpolating from the particle positions. The interpolation takes the
form of a simple summation over the particles with each term weighted
according to distance from the point in question. The weighting
function is known as the interpolation kernel. For example, the
interpolated value for the
density at some point ${\bf r}$ in the fluid 
\begin{equation}
\rho ({\bf r}) = \sum_i^n m_i\,W({\bf r}-{\bf r}_i,h),
\end{equation}
where the sum is over all the particles. Here, $m_i$ and ${\bf r}$ are  the
mass and position of particle $i$ respectively. $W$ is the
interpolation kernel which has a characteristic length-scale $h$,
commonly called the smoothing length. For a general introduction to SPH the
reader is referred to Monaghan (1992).

In Murray (1996) an SPH code specifically modified for accretion disc
problems was described. The key feature of this code was the use of an
artificial viscosity term in the SPH equations to represent the shear
viscosity $\nu$ known to be present in observed discs.
The artificial viscosity term introduces, in the continuum limit, a
fixed  combination of shear and bulk viscosities to the fluid. The
viscous force per unit mass,
\begin{equation}
{\bf a_{\rm v}}= \kappa\, \zeta\, c\, L (\nabla^2 {\bf v} + 2 \nabla ({\bf
\nabla} \cdot {\bf v})).
\end{equation}
$\kappa$ is an analytically determined constant that depends upon the
 kernel. $\kappa=1/8 $ and $1/10$ for the cubic spline kernel in two
 and three dimensions respectively. $c$ is the sound speed, ${\bf v}$
 is the fluid velocity and $L$ a viscous length scale. In previous
 work $L$ was taken to be equal to the smoothing length $h$. Here
 however we relax that constraint.
 $\zeta$ is a
 dimensionless parameter. Note that in most SPH papers this parameter
 is denoted $\alpha$ but we have followed Murray (1996) and renamed it
 to avoid a confusion of subscripts.

In the interior of Keplerian discs
 we can neglect the velocity divergence and we see that the artificial
 viscosity term generates a shear viscosity
\begin{equation}
\nu=\kappa\,\zeta\, c\, L.
\end{equation}
We can control the shear viscosity throughout the disc by modifying
$\zeta$ and $L$, and so obtain a functional form very similar to the
Shakura-Sunyaev form used in accretion disc theory.

Several tests
of this code were presented in Murray (1996).
The code has been used
to look at tidally unstable discs (Murray 1998, 2000), tilted discs
(Murray \& Armitage 1998) and counter-rotating discs (Murray, deKool
\& Li, 1999) around accreting pulsars. Kornet \&
R\'o\.zycka (2000), using an Eulerian code (quite distinct algorithmically
 from the SPH code we use), have reproduced several of the results of Murray (1998).

\subsection{Outbursts}

As mentioned in the previous section, quiescence is associated with a
very low value of the Shakura-Sunyaev shear viscosity parameter
($\alpha$) but also a low temperature. In outburst, angular momentum transport
is much more rapid because the temperature is much higher and $\alpha$
is much larger.
In previous papers (Murray 1998, Armitage \& Murray 1998) preliminary
attempts were made to model outbursts by instantaneously increasing
the value of the shear viscosity throughout the entire disc. 
Such an approach only enabled us to model the most basic features of
an outburst, 
and we could not of course follow the propagation of state changes through the disc.
The
principal modification to the code made for this work was to allow the
viscosity to change locally in response to disc conditions. This was
easy to do, simply requiring that each particle carry a variable $\zeta$
that determined its 'viscosity'. To determine the viscosity of the
interaction of any particular pair of particles we use the harmonic
mean of the two $\zeta$ values. Cleary \& Monaghan (1999) found that such a
form gave good results for heat conduction between materials with
vastly different properties.

\begin{figure}
\psfig{file=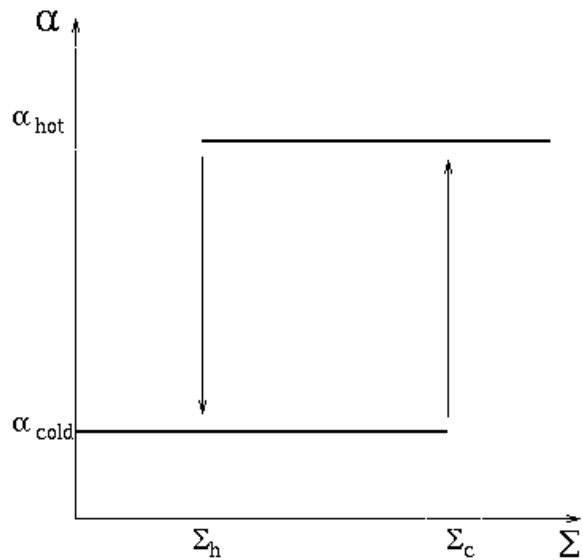,width=8cm}
\caption{The simplified surface density trigger conditions used in the
code. $\Sigma_{\rm c}$ and $\Sigma_{\rm h}$ are functions of 
radius in the disc.}
\label{dtrig}
\end{figure}

All that remained was to determine
how changes in viscosity were  to be triggered.
The simplest approach is to let the  shear viscosity of each SPH
particle be
determined by the local surface density $\Sigma$. If, for some region
of the disc in the quiescent state, the surface
density is less than some critical value $\Sigma_{\rm c}$ that region
of the disc is assumed to be stably quiescent, and the particle's shear
viscosity remains at a value appropriate for a cool disc. 
Should $\Sigma$ increase to be greater than
$\Sigma_{\rm c}$ then we consider that region of the disc to have been
`triggered' into the hot state. We then let the particle's shear viscosity
increase to a value appropriate for a hot disc (the details of the
transition are described below). The viscosity continues adjusting to
its new value
even if $\Sigma$ subsequently drops below $\Sigma_{\rm c}$. 
Conversely a hot region of the disc is stable as long as
its surface density remains above a second critical value $\Sigma_{\rm
h} < \Sigma_{\rm c}$. However if a portion of the disc  
has $\Sigma < \Sigma_{\rm h}$ then that particle's 
shear viscosity parameter will reduce to the quiescent value
(See figure~\ref{dtrig}).  $\Sigma_{\rm h}$ and $\Sigma_{\rm c}$
correspond to $\Sigma_{\rm{min}}$ and $\Sigma_{\rm{max}}$. We preserve
their radial dependence but reduce the magnitude of their gradients to
reduce the run-time of the code. A full analysis of the scaling is
given in Section 3.2.

In previous work (Murray 1998,
Armitage \& Murray 1998) $\alpha$ was set to change
instantaneously between quiescent to outburst values.
In this work, we assume that when the instability is triggered, the viscosity
changes on the thermal time scale. 
As we have limited knowledge of the mechanism responsible for
accretion discs' high viscosity, we can only propose an appropriate
functional form for the transition. We use the hyperbolic tangent
function as it allows us to capture both the initial exponential
change in $\nu$ and a smooth asymptotic approach to its final
value. In terms of the Shakura-Sunyaev viscosity parameter,
the equation for the transition from quiescence ($\alpha=\alpha_{\rm cold}$) to
outburst ($\alpha=\alpha_{\rm hot}$) is
\begin{equation}
\alpha(t)= \frac{(\alpha_{\rm hot}+\alpha_{\rm cold})}{2} + 
\frac{(\alpha_{\rm hot}-\alpha_{\rm cold})}{2} \tanh (\frac{t}{t_{\rm th}} - \pi)
\end{equation}
and the converse, for the transition from $\alpha_{\rm hot}$ to
$\alpha_{\rm cold}$ is
\begin{equation}
\alpha(t)= \frac{(\alpha_{\rm hot}+\alpha_{\rm cold})}{2} - 
\frac{(\alpha_{\rm hot}-\alpha_{\rm cold})}{2} \tanh (\frac{t}{t_{\rm th}} - \pi).
\end{equation}

\begin{figure}
\psfig{file=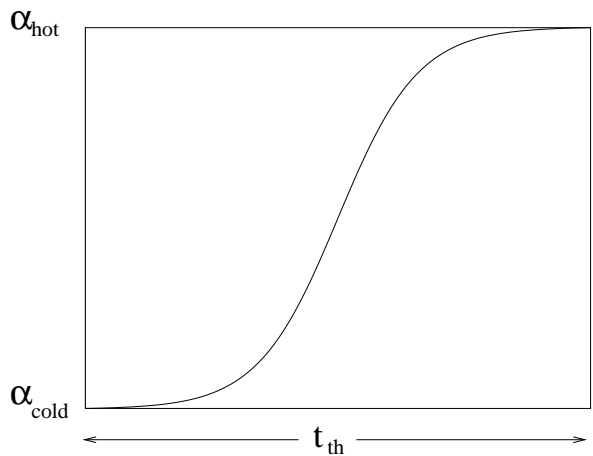,width=8cm}
\caption{Functional form of the outburst trigger. The viscosity is
switched on a total time-scale $\rm t_{th}$.}
\label{trig}
\end{figure}

The functional form of the trigger is shown in Figure~{\ref{trig}}
Of course, $\tanh$ doesnt have compact support, so as soon as a
transition has been triggered we add a small artificial offset
to $\alpha$.

In the following section we describe simulations completed with the
new variable viscosity. As well as using a density trigger in which
single SPH particles could change state, we implemented a more coarse
grained approach. The disc was divided into a set of concentric
annuli (typically 100 were used). 
When the mean surface density in a given annulus met the
triggering conditions, then all particles in that annulus changed
state (again on the   thermal time scale). This intermediate,
azimuthally smoothed,  approach
improved the speed of the code by $\sim$ 50\%, and provides a suitable basis for
incorporation of a full thermodynamic treatment of the thermal
instability. The calculations described in the next section
demonstrate that these two approaches to the viscosity switching are
consistent with one another.

As it is written, the triggering routine can simply be replaced by a
more thermodynamically sophisticated  routine. Preliminary calculations
have indeed been made. However, as these simulations are proving more computationally
demanding, we leave them to a later paper.

\section{Results}

We show that our fast azimuthally-smoothed approximation gives a good
representation of the physical behaviour of the system and move on to
present a wide range of results and analysis for a simulation of the
dwarf nova SS Cygni. SS Cygni is a very well-observed dwarf nova of
the U Gem type. It has an orbital period of 0.275130 days and a mass
ratio (defined as mass of the mass-losing secondary to that of the
accreting white dwarf primary) of 0.59 $\pm$ 0.02, with $M_{1}$ = 1.19
$\pm$ 0.02 $\rm{M_{\odot}}$ (Friend et al, 1990).

All the results presented in this paper are for {\em steady state}
discs, which we define to be those which return to the same quiescent
level between outbursts. We take a constant mass-transfer rate of
$10^{-9} \rm{M_{\odot} yr^{-1}}$ throughout the simulations, following
Cannizzo (1993b).

\subsection{Comparison of triggering regimes}

The simulation was performed with both local triggering and
azimuthally smoothed triggering as detailed above.
In the case of local triggering, two regimes are contrasted;
(i)$\tau_{\rmn{trigger}}$ = 250 s,
(ii)$\tau_{\rmn{trigger}}$ = 5000 s.
For the azimuthally smoothed case, we choose a long trigger timescale
of 7500 s.
The thermal time-scale $t_{\rmn{th}}$ and the dynamical (Kepler)
time-scale $t_{\rmn{\phi}}$ are related by
 
\begin{equation}
~~~~~~~~~~~~~~~~~~~~~~~~~~~~~~~~t_{\rmn{\phi}}\sim\alpha t_{\rmn{th}},
\end{equation}

\noindent
SS Cygni has
an orbital period of almost 400 minutes and a Kepler timescale of
around 250 seconds at the circularisation radius. This is the radius
at which the gas has the same specific angular momentum as it had on
passing through the $L_1$ point (i.e. as it is injected into the
model). It is the radius at which infalling gas would first orbit the
primary before formation of an accretion disc. Hence for
realistic values of $\alpha~ (<< 1.0)$, $t_{\rmn{th}}$ will be of the
order of a few thousand seconds, which we use in the azimuthally
smoothed code and case(ii).

\begin{figure}
\psfig{file=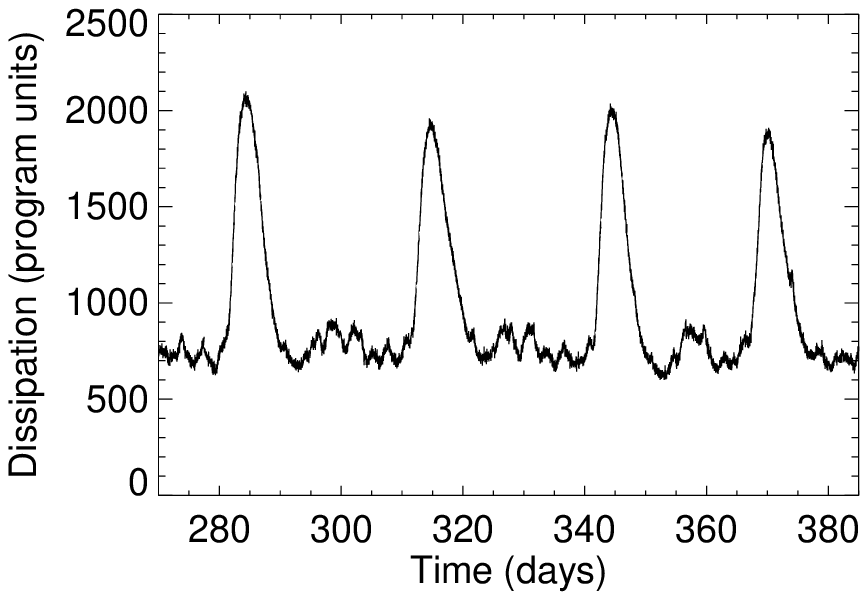,width=8cm}
\psfig{file=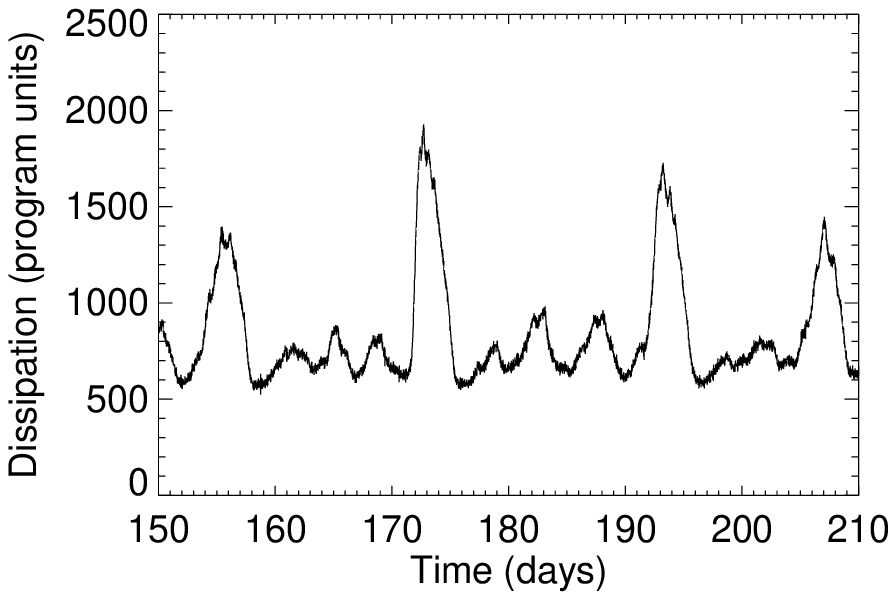,width=8cm}
\psfig{file=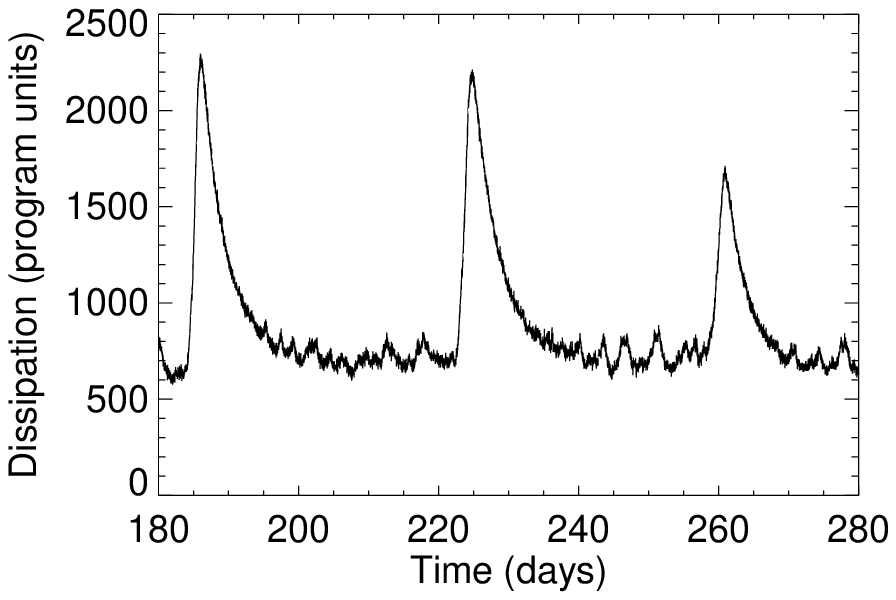,width=8cm}
\caption{Top : Light curve (total viscous dissipation) in
SS Cygni simulation with azimuthal smoothing and
$\tau_{\rmn{trigger}}$ = 7500s. Middle : Light curve
with local triggering and $\tau_{\rmn{trigger}}$ = 250 s. Bottom : Light
curve with local triggering and $\tau_{\rmn{trigger}}$ = 5000 s}

\label{diss}
\end{figure}

Figure~\ref{diss} contains plots of total viscous dissipation against
time for the different regimes. The outbursts in both cases are
coherent, that is a large region of the disc is transformed to the hot
state. It is clear that the azimuthally smoothed result is 
consistent with a local trigger time-scale somewhere between these two
cases. The peak-to-peak time-scale between outbursts is around 30 days
in the smoothed code, 20 days in the shorter timescale code and 40
days in the longer time-scale code. Our fast azimuthally smoothed code is consistent with
the physically realistic case of a  thermal time-scale greater than the
Kepler time-scale.

\subsection{Light curves}

\begin{figure*}
\vbox to 160mm {\psfig{file=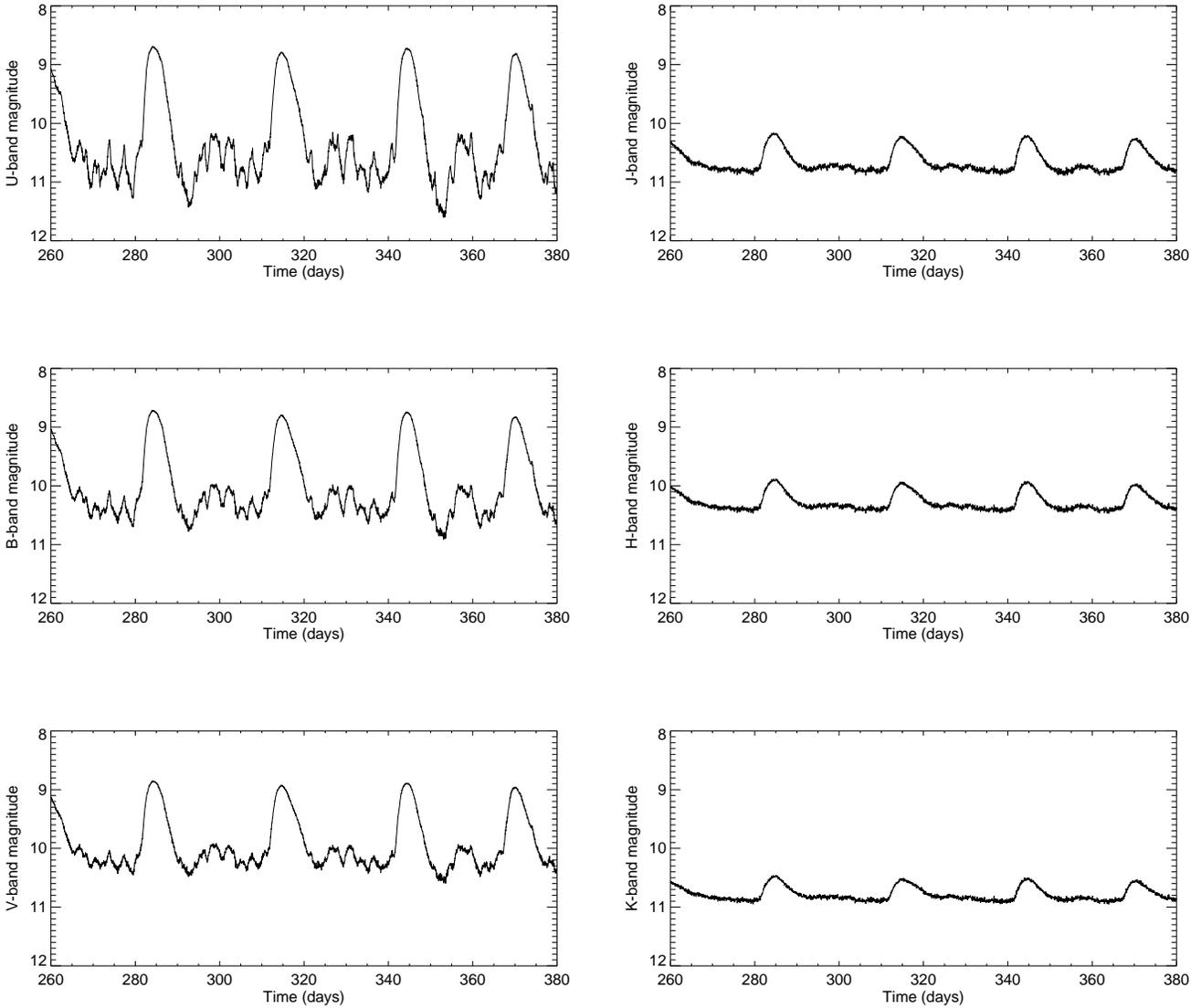,width=18cm}}
\caption{U,B,V,J,H and K filter band light
curves (azimuthally-smoothed model). The light
curves show three distinct states - outburst, quiescence and minioutbursts}
\label{ubv}
\end{figure*}

Figure~\ref{ubv} shows light curves calculated from the SS Cygni simulation in
the U,B,V,J,H and K bands. These are calculated by assuming that the
disc is optically thick and each annulus of gas in the disc radiates as a black body. We
then simply integrate the Planck function over different wavebands and
sum over the annuli to yield the light curves. 

The light curves show regular normal outbursts with a comparatively rapid
rise to maximum and a slightly longer fall to quiescence. The
amplitude of the normal outbursts in the V band varies around 1.5
magnitudes on our diagram, with a rise-time of $\sim$ 2.5 days and a
decay-time of $\sim$ 4.5 days. SS Cygni is observed to brighten from 12th
magnitude in quiescence to 8th magnitude in outburst. Our result is
suppressed by the artificially high value of the Shakura-Sunyaev alpha
parameter in the two states, which we use to reduce the run-time of
the code. We can make an estimate of the compression factor of the
amplitude as follows. The dissipation rate of an accretion disc D(R) is
directly proportional to the torque exerted by one annulus on another
G(R) :

\begin{equation}
   D(R) = {G(R)\over{4 \pi R}}{d\Omega\over{dR}}
\end{equation} 
\noindent
where $\Omega$ is the angular velocity of the gas in the annulus. The
viscous torque is 

\begin{equation}
   G(R) = 2 \pi R\nu\Sigma{R^{2}}{d\Omega\over{dR}}
\end{equation}
\noindent
where $\nu$ is the viscosity, so

\begin{equation}
   D(R) ={1\over2}\nu\Sigma(R{d\Omega\over{dR}})^{2}
\end{equation}
\noindent
The dissipation rate is therefore proportional to $\nu\Sigma$. Equation 7
gives the relationship between $\nu$ and the alpha-parameter ($\zeta$
in Eq.7). In an isothermal disc, the ratio of total dissipation in
outburst to that in quiescence is

\begin{equation}
   {{D_{\rm{outburst}}}\over{D_{\rm{quiescence}}}} = 
{(\alpha\Sigma)_{\rmn{outburst}}\over(\alpha\Sigma)_{\rmn{quiescence}}}.
\end{equation}
\noindent 
We use a simple 1:10 ratio of $\alpha$ irrespective of
surface density $\Sigma$. The density triggers are set at

\begin{equation}
   \Sigma_{\rm{max}} = 16.67{R\over{a}} ~\rm{gcm^{-2}}
\end{equation}
\begin{equation}
   \Sigma_{\rm{min}} =  6.250{R\over{a}} ~\rm{gcm^{-2}}
\end{equation}
\noindent 
where a is the binary separation.
The surface density at the peak of outburst will be close to
$\Sigma_{\rm{min}}$, while the surface density in quiescence just
before an outburst will be close to $\Sigma_{\rm{max}}$. This is
clearly shown in Fig.~\ref{dens}. Therefore, in our isothermal disc
simulation,

\begin{equation}
{{D_{\rm{outburst}}}\over{D_{\rm{quiescence}}}} = 
{{\alpha_{\rm{hot}}\Sigma_{\rm{min}}}\over{\alpha_{\rm{cold}}\Sigma_{\rm{max}}}}
\simeq 3.7.
\end{equation}
\noindent    
Real discs, however, are not isothermal. Rewriting Eq.7 in terms of
more familiar quantities for a thin Shakura-Sunyaev disc,
\begin{equation}
\nu = \alpha c_{\rm{s}}H \sim \alpha c_{\rm{s}}^{2}({{R^{3}}\over{GM}})^{1/2}
\end{equation}
\noindent 
where $c_{\rm{s}}$ is the sound speed and H is the scale height. In
the limit where we can neglect radiation pressure,
\begin{equation}
c_{\rm{s}}^{2} = {P\over\rho} \propto T_{\rm{c}}
\end{equation}
\noindent 
where $T_{\rm{c}}$ is the midplane temperature. The critical surface densities
will be given by (3) and (4) and in a real disc at fixed radius, 

\begin{equation}
{D_{\rmn{outburst}}\over{D_{\rmn{quiescence}}}} =
{8.25{\alpha_{\rmn{hot}}{T_{\rm{c,hot}}}{{\alpha_{\rmn{hot}}^{-0.8}}}}\over{11.4\alpha_{\rmn{cold}}{T_{\rm{c,cold}}}{{\alpha_{\rmn{cold}}^{-0.86}}}}}
\end{equation}
\noindent
Using Cannizzo's (1993b) values of $T_{\rm{c,hot}}\simeq$ 60,000K and
$T_{\rm{c,cold}}\simeq$ 4,000K, and setting
$\alpha_{\rmn{cold}}$=0.01 and $\alpha_{\rmn{hot}}$=0.1, this ratio is about
13.1.
The compression factor introduced in dissipation amplitude by the code
is about 3.5, bringing our result closer to the eqivalent observed
amplitude in SS Cygni.

In the quiescent phase, there appears to be an additional modulation
in luminosity. These aperiodic {\em minioutbursts} appear in clusters
and are particularly blue, suggesting  that only the hot inner portion
of the disc is involved in producing the additional luminosity. These
variations are indeed observed in the V-band light curve of SS Cygni
(see, for example, the AAVSO light curves).

\begin{figure}
\psfig{file=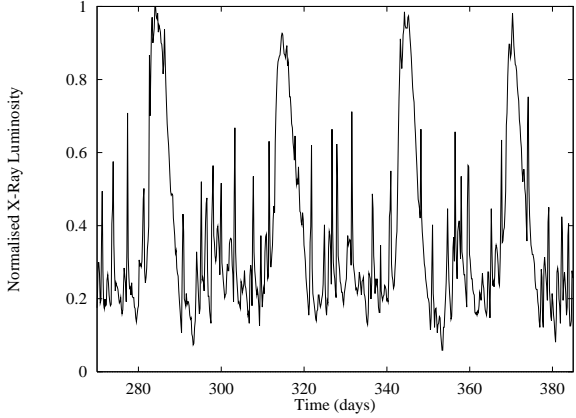,width=8cm}
\caption{Top : X-ray light curve generated from number of simulation
particles accreted onto the primary.}
\label{xray}
\end{figure}

It is possible to correlate the number of particles being accreted
onto the white dwarf primary with an expected X-ray luminosity for the
system by simple consideration of gravitational energy released. For
simulation particles of mass $M_{\rmn{p}}$, the X-ray luminosity
produced is
\begin{equation}
    L_{\rmn{X}}=\epsilon{{GM_{\rmn{p}}}\over{R}}{dM\over{dt}}
\end{equation}
where $\epsilon$ is an unknown efficiency factor expected to be $<< 1$.
\begin{figure}
\psfig{file=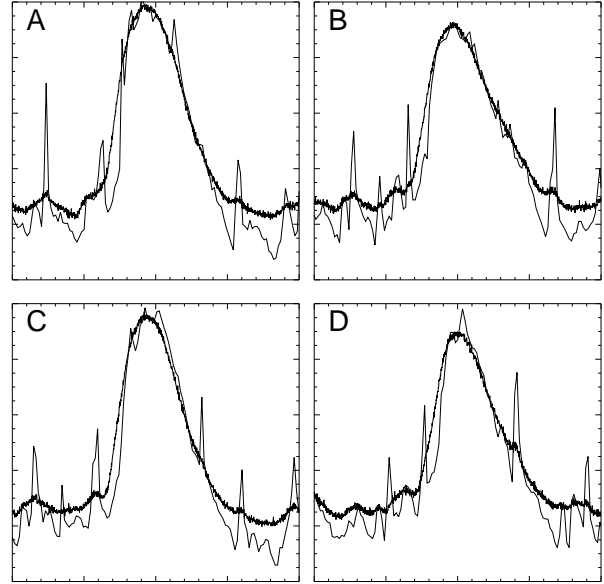,width=8cm}
\caption{Overlay of V-band (dark line) and X-ray emission for the four
outbursts in Figure~\ref{ubv}. Each plot has been normalised to the
peak of curve A and shows a 20-day time series.}
\label{comp}
\end{figure}
An X-ray light curve can be built up in this way and is
presented in Figure~\ref{xray} for our simulation. In quiescence and
on the decline from outburst, the X-ray emission follows the optical
emission very closely; indeed our X-rays appear to be very sensitive
to fluctuations in the V-band in quiescence. However, there is a
noticeable and significant delay in the onset of the outbursts. This
is shown in Figure~\ref{comp} for each of the four outbursts. Each
plot shows 20 days; referring back to Figure~\ref{ubv} we have
A : 275-295 days, B : 305-325 days, C : 335-355 and D : 360-380 days.
The optical rise leads the X-ray rise by up to a day in all our
outbursts. This can be interpreted as the time required for the gas in
the part of the disc which initially goes into outburst to migrate
through the inner disc before it is accreted onto the primary. The
measurement of the delay is in good agreement with observation; Jones $\&$
Watson (1992) found a typical delay of 0.5 - 1.1 days from EXOSAT
observations. The rise-time is also seen to be much faster in X-rays;
we find a rise-time of $\sim$12 hours in X-ray and $\sim$2.5 days in
the optical. The climb in V-band emission is a gradual process as more
and more gas in the disc is transformed to the high state, whereas the
onset of enhanced accretion onto the white dwarf itself is
instantaneous, triggered when the additional material from the
outbursting disc arrives at its surface.

\subsection{Disc Analysis}

\begin{figure*}
\vbox to 220mm {~~~~~~~~~~~~~\psfig{file=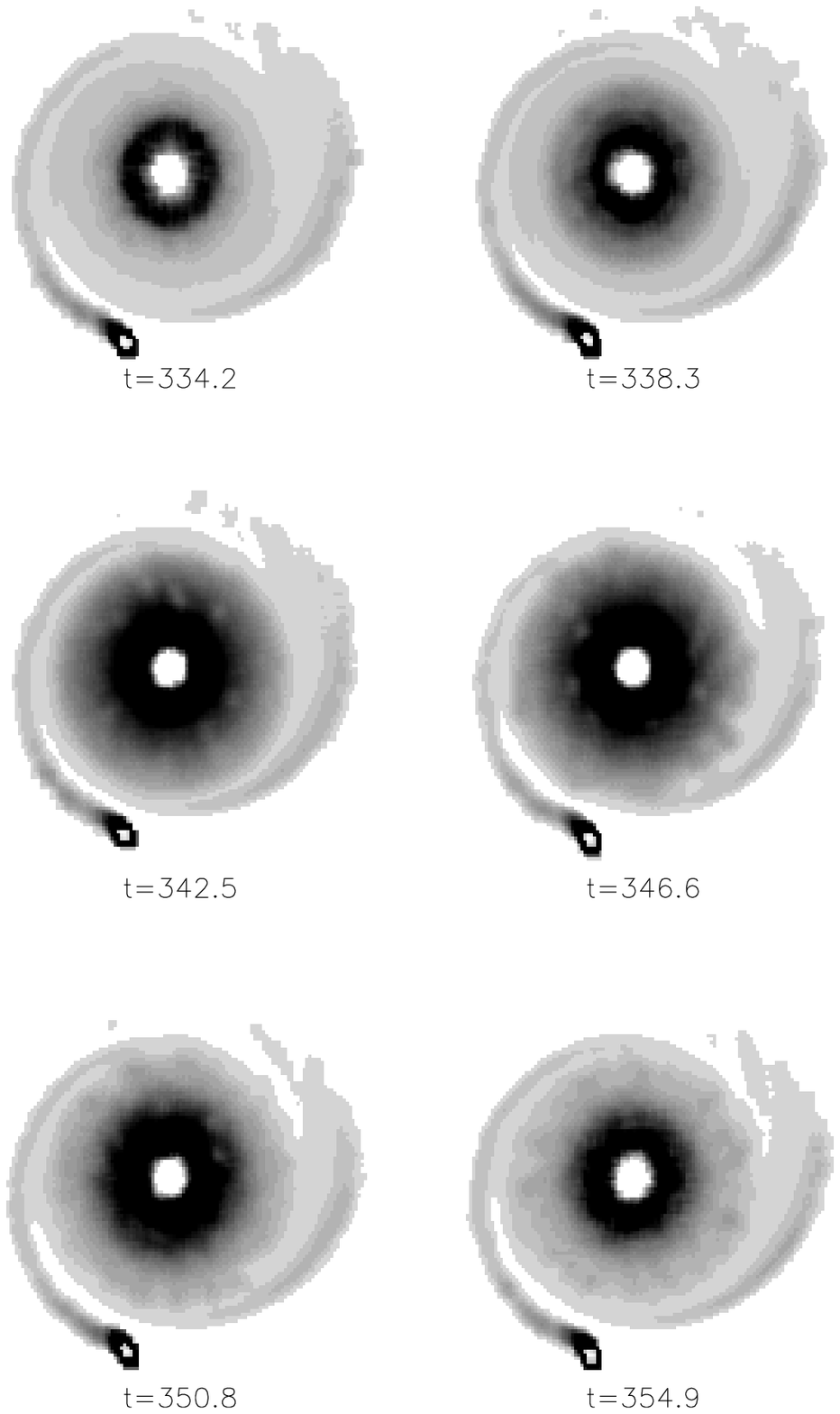,width=12cm}
\\
~~~~~~~~~~~~\psfig{file=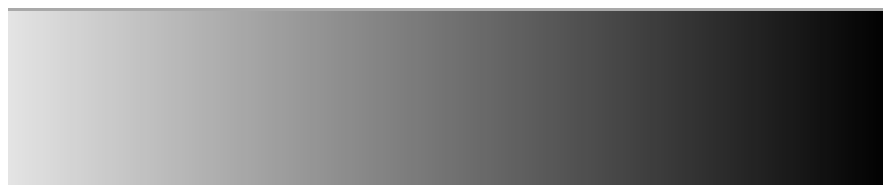,width=6cm}}
\caption{Evolution of total viscous dissipation in the disc through an
outburst. The greyscale is logarithmic with white at $10^{7} \rm{ergs^{-1}cm^{-2}}$
through to black at $10^{9} \rm{ergs^{-1}cm^{-2}}$. The discs are
aligned such that the primary-secondary axis is vertical and the
hot-spot appears at the bottom of each frame. Spiral shock arms are
also clearly evident. t is the time in days. Note that the central
part of the disc (white) is not modelled in the simulation.} 
\label{discs}
\end{figure*} 

\begin{figure}
\psfig{file=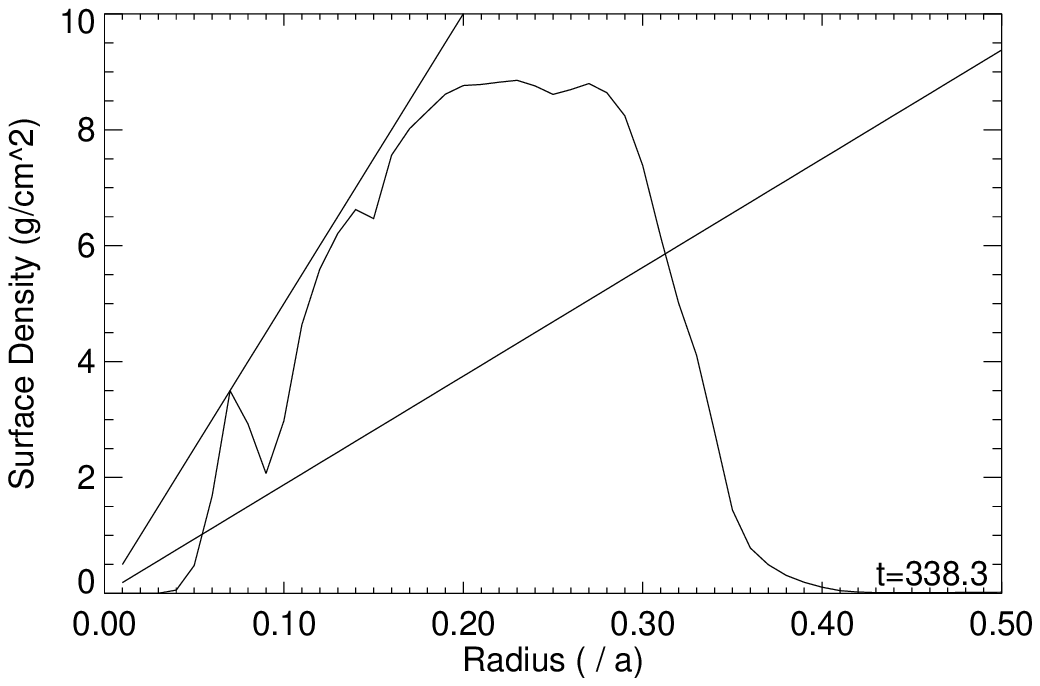,width=8cm}
\psfig{file=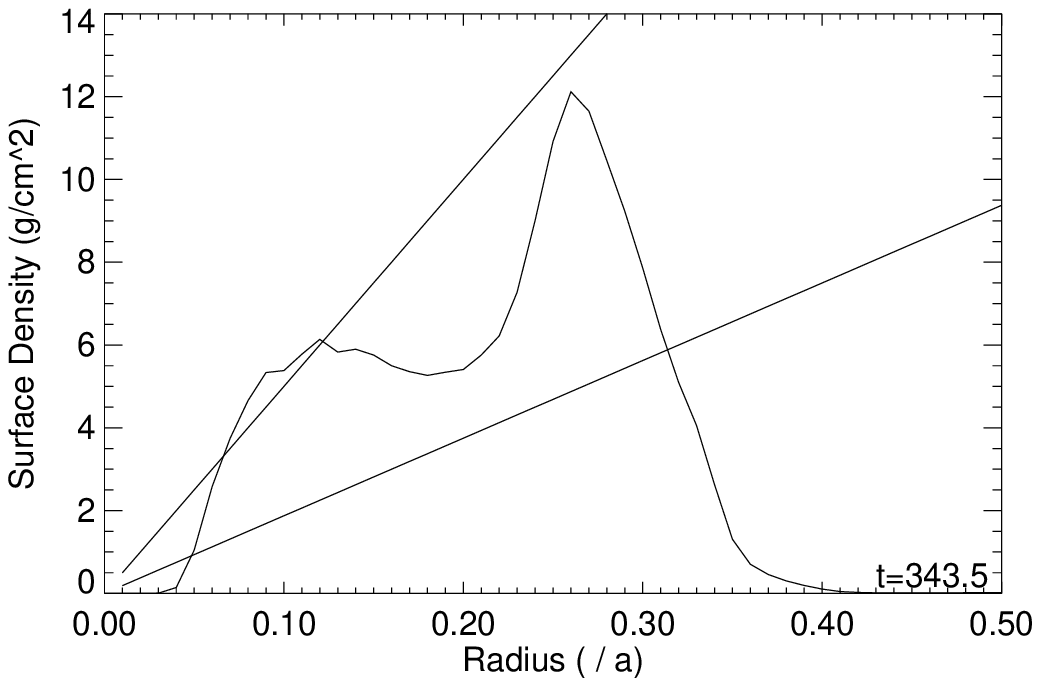,width=8cm}
\psfig{file=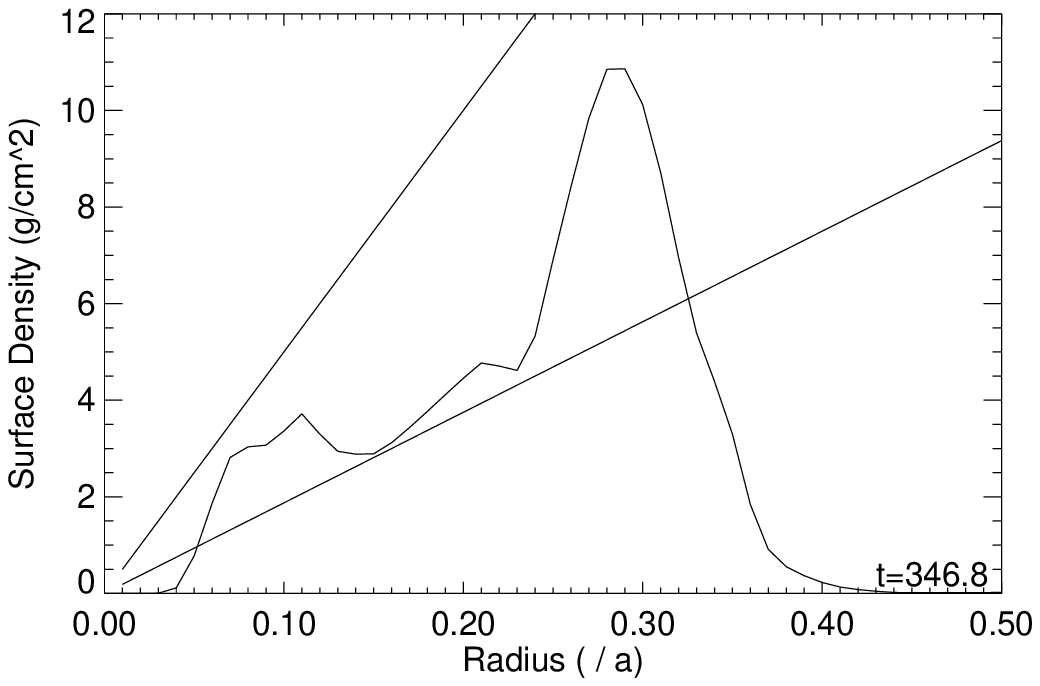,width=8cm}
\caption{Evolution of azimuthally-averaged surface density of the disc
through an outburst. The straight lines show the radial dependence of the upper
and lower triggers. a is the binary separation, 1.8 x $10^{11}$cm
in SS Cyg.}
\label{dens}
\end{figure}

In this section we follow the evolution of an accretion disc through
an outburst cycle.
The light curves give much information about the physical processes
taking place in the disc during quiescence and outburst, but a much
clearer representation is obtained by following the dynamical
behaviour of the accretion disc itself. Figure~\ref{discs} is a montage of
images of the accretion disc through a normal outburst.
We find that even in quiescence (t=334.2), a small fraction of the
inner disc remains permanently locked in the hot state. The linear
relation between the critical surface density triggers and the disc
radius ensures that particles in the inner disc are continually
cycling between the low and high states (the triggers are
very close together here and also very low - these conditions are
easily satisfied by particles close to the primary).
The rise to outburst is rapid and is intitiated in the inner part of
the disc. A density wave (directly analagous to a heating wave in a
full thermodynamic treatment) moves both outwards and inwards,
transforming a large fraction of the disc to the hot, high viscosity
state. Subsequently the disc falls (less rapidly) back into quiescence 
by another density wave (analagous to a cooling wave). This pattern is repeated for all the normal
outbursts. Approximately the same fraction of the disc is seen to
reach the high state in each outburst. The maximum extent of the
hot portion is seen to vary to a small degree, but appears ultimately
to be 
limited to a radius near the position of the spiral shock arms. This can be readily observed
in the light curve - some outbursts reach a slightly higher peak
luminosity than others, although the maximum variation is no more than
a few percent.

The outburst is, of course, triggered by the local surface density in
the disc and as the gas in the disc is accreted inwards on a viscous
timescale there is a corresponding change in that local surface
density. This response is shown in Figure~\ref{dens}. The disc clearly
drains during the outburst phase until the local surface density
reaches the lower trigger level, after which the outburst dies away
and the disc is replenished. These density profiles are extremely
similar to those found by Stehle (1999) in a full thermodynamic
one-dimensional treatment (not SPH).

\begin{figure}
\psfig{file=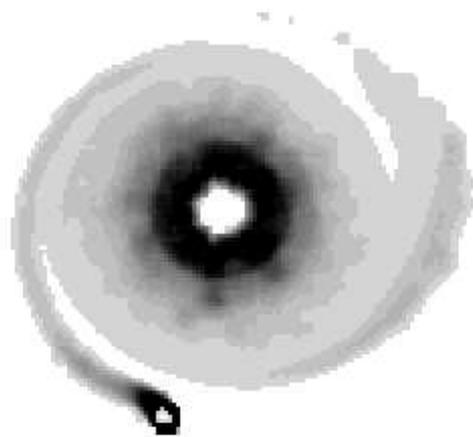,width=8cm}
\caption{The disc in minioutburst. The outburst has
not propagated far enough to become a normal outburst. The greyscale
is the same as in Fig.~\ref{discs}.}
\label{mini}
\end{figure}

Figure~\ref{mini} shows a disc in minioutburst, and confirms our observation
from the light curve in section 3.2 that it is the inner disk which is
hot in a minioutburst. The outburst is initiated very near the inner
edge of the disc and never propagates very far. The reason for this
can be seen in Figure~\ref{dens}. Just before a normal outburst (top
panel), the surface density profile follows the line of the upper
trigger very closely. Any part of the disc that goes into outburst
will lead to a coherent normal outburst throughtout most of the
disc. After a normal outburst (bottom panel), the only part of the
disc near the upper trigger is near the disc centre. All the remainder
of the disc has been drained. Hence while the centre of the disc can
go into outburst, the majority of the disc cannot. This is what
happens in a minioutburst.

\begin{figure}
\psfig{file=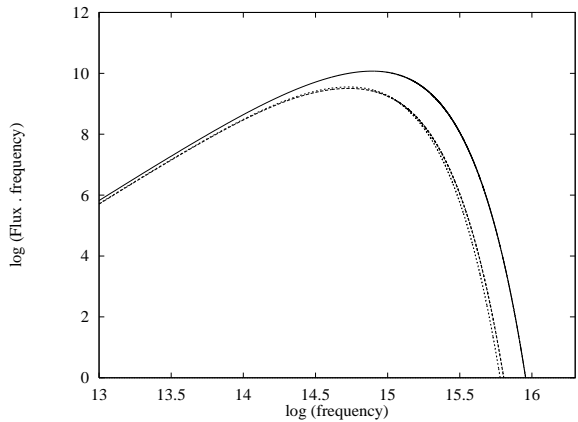,width=8cm}
\caption{Disc spectra in quiescence (bottom curve), minioutburst (middle) and
outburst (top). The disc in minioutburst is bluer than the quiescent disc.}
\label{spec}
\end{figure}

Disc spectra, temperature profiles and luminosity profiles clearly
show the contrast in the physical properties of the disc in the
different states. The spectra, in Figure~\ref{spec}, have been computed for the
disc in outburst, quiescence and minioutburst. The blue, hot inner
disk behaviour in minioutburst becomes clear in comparison with the
quiescent spectrum.

\begin{figure}
\psfig{file=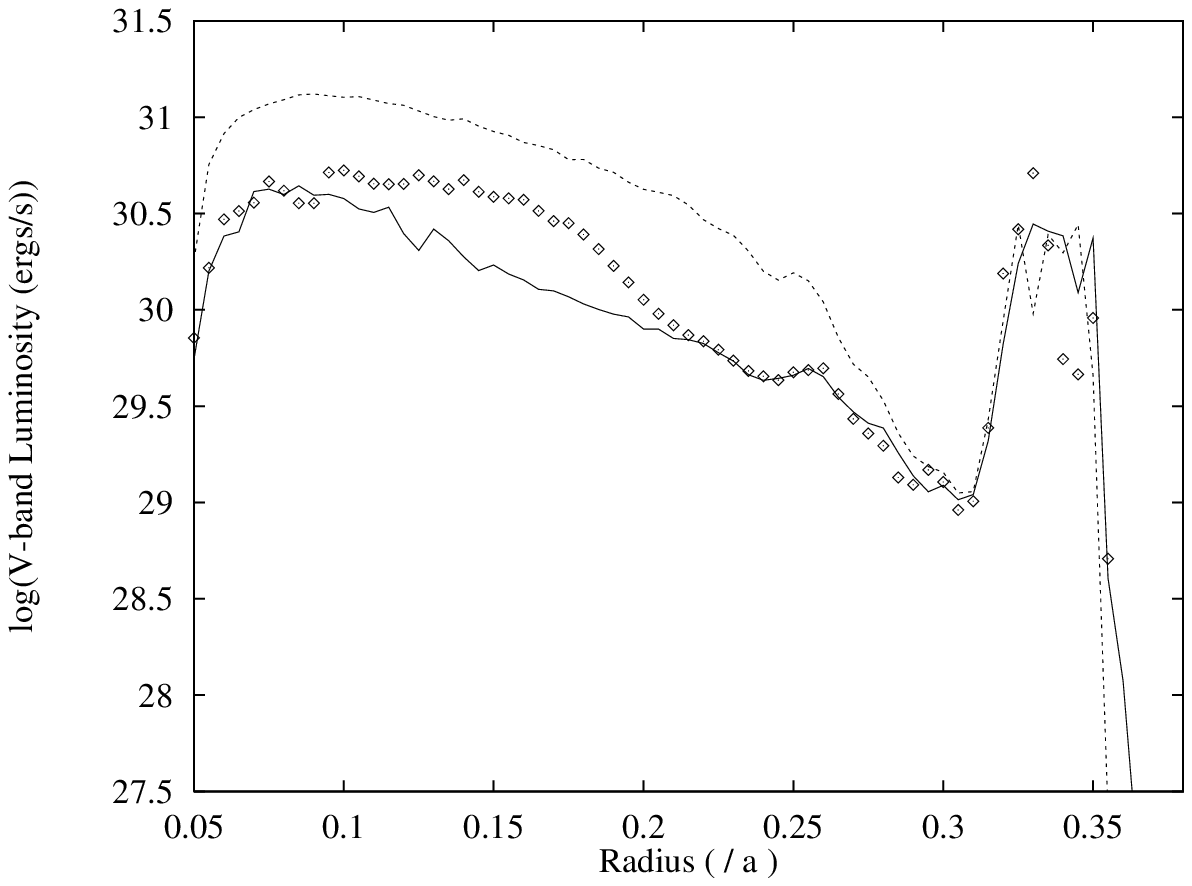,width=8cm}
\psfig{file=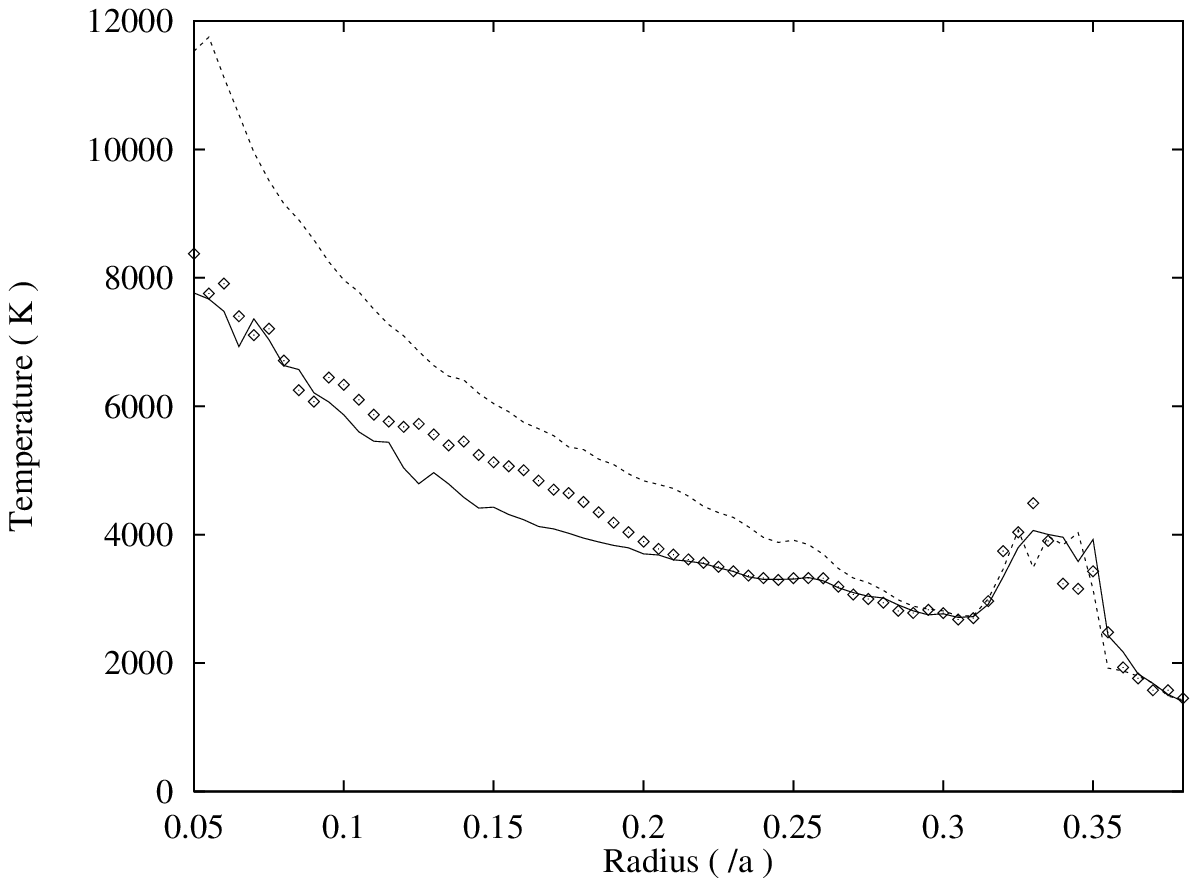,width=8cm}
\caption{V-band luminosity and temperature of the disc
evolving through the rise to normal outburst. The inner region of the
disc remains in outburst even during quiescence (solid line) and the normal
outburst is initiated at a radius around $0.15a$ (points). The dashed
line represents the profile at peak outburst.}
\label{prof}
\end{figure}

The luminosity and temperature profiles in Figure~\ref{prof} have been
calculated on the rise to a normal outburst, and correspond to the
discs in the first three frames of Figure~\ref{discs}. All the profiles show an
increased region of viscous dissipation (and temperature) near the
outer edge of the disk, where the surface density peaks as matter
enters the disc from the accretion stream; this is the {\em{hot-spot}}.
This normal outburst is apparently triggered at a radius of
approximately 0.15a (near the circularisation radius - 0.11a in SS
Cyg) and the heating wave is seen to propagate in both 
directions. We find that in quiescence the disc is typically
approximately isothermal at radii greater than 0.15a at temperatures
around 4000 K. In outburst we find temperatures ranging from 6000 K to
12000 K in the inner half of the disc and 4000 K to 6000 K in the
outer half. Bobinger et al (1997) have perfomed eclipse mapping of the
dwarf nova IP Peg , which has strikingly similar parameters to SS Cyg
($q=0.58, P=0.158 d$), on the decline from outburst. They found an
inner disc at 7000 K to 9000 K and an outer disc with temperatures
declining to a quiescent level of 3000 K to 4000 K. Our simulated
results are in good agreement with these
observations. Figure~\ref{prof} also shows that the hot-spot is
consistently around 1500 K hotter than the surrounding regions, but
the increase in luminosity from these regions is much
more marked - the viscous dissipation is 30 times higher here than in the
part of the disk just inside them. The accretion rate from the
secondary is constant in this simulation, so we find
a constant hot-spot temperature. It is also interesting to note,
however, that the luminosity generated in the hot-spot of the
quiescent disc is comparable to that generated by the very hot inner
part of the disc. 

\section{Discussion}

We have performed the first two-dimensional treatment of dwarf nova
outbursts. The method gives light curves in different wavelength
bands, disc spectra,density and temperature profiles which all agree well with
observations and full thermodynamic analyses.

The advantages of using a two-dimensional code to model dwarf nova
outbursts are immediately obvious. Tidal forces are not approximated
as in one dimensional codes and therefore much more detail is revealed in the
results of these current simulations. One dimensional simulations
have always predicted extremely regular outburst behaviour, both in
recurrence pattern and outburst profile.  There is now much more scope
for exploring the behaviour of these systems; the simulated
observational characteristics presented here are just a selection of
the possible areas which can be explored in future work.

We have applied these methods successfully to a study of the SU Ursae Majoris
systems (Truss et al, 2000) in addition to
the dwarf novae with less extreme mass ratios considered here. A
two-dimensional code is ideal for exploring tidal effects and
instabilities in such systems. 

Observations of dwarf novae are by no means comprehensive or
continuous, but this situation should be remedied in the near future
with more telescope time dedicated to them. In consequence, it will be
possible to test several of the results in this paper. Although our
knowledge of the behaviour of dwarf novae in the V-band is good thanks to the
efforts of organisations such as the AAVSO, there is little data available in
other bands, so new observations will be able to probe the inner
regions of the disc and (via X-ray observations) boundary layer. Our
findings relating to a hot inner disc and the occurrence of
minioutbursts should have some testable observational consequences.

The next stage of development of any numerical model such as this should
be twofold : the incorporation of full thermodynamics into the
simulation and the extension to three dimensions. This is not a
difficult task with the present scheme, just a lengthy one in
computational terms. The increasing availability of parallel machines
and improvements in CPU perfomance should make these refinements a
workable proposition. 

\section*{Acknowledgments}
Research in theoretical astrophysics at the University of Leicester is
supported by a PPARC rolling grant. Many of the calculations were performed using GRAND, a high
performance computing facility funded by PPARC and based at the
University of Leicester. The authors are grateful to Dr R. Stehle,
Prof. J. Pringle, Prof. A. King and an anonymous referee for helpful 
discussions or comments. MRT acknowledges a
PPARC research studentship and a William Edwards Charitable Trust
bursary for postgraduate study.

\end{document}